\documentclass[11pt]{llncs}
\usepackage{latexsym}
\usepackage{amssymb}
\begin{document}
\newcommand{\eat}[1]{}
\title{Inference of termination conditions for numerical loops}
\author{Alexander Serebrenik\thanks{supported by GOA: ``${LP}^{+}$: a second gen
eration
logic programming language''.}, Danny De Schreye}
\institute{Department of Computer Science, K.U. Leuven\\
Celestijnenlaan 200A, B-3001, Heverlee,
Belgium\\E-mail: Alexander.Serebrenik@cs.kuleuven.ac.be}
\maketitle

\begin{abstract}
We present a new approach to termination analysis of numerical
computations in logic programs. Traditional approaches fail to
analyse them due to non well-foundedness of
the integers. We present a technique that allows to overcome these
difficulties. Our approach is based on transforming a program in way 
that allows integrating and extending techniques 
originally developed for analysis of 
numerical computations in the framework of query-mapping pairs 
with the well-known framework of acceptability.
Such an integration not only contributes to the understanding of 
termination behaviour of numerical computations, but also
allows to perform a correct analysis of such computations automatically,
thus, extending previous work on a constraints-based approach to termination.
In the last section of the paper we discuss possible extensions of 
the technique, including incorporating general term orderings.
\end{abstract}

\section{Introduction}
Numerical computations form an essential part of almost any 
real-world program. Clearly, in order for a termination analyser to be of
practical use it should contain a mechanism for inferring termination
of such computations. However, this topic attracted less attention of
the research community. In this work we concentrate on automatic
termination inference for logic programs depending on numerical computations.
Dershowitz {\em et al.}~\cite{Dershowitz:Lindenstrauss:Sagiv:Serebrenik}
showed that termination of general numerical computations,
for instance on floating point numbers, may be 
contr-intuitive, i.e., the actually observed behaviour does not necessary
coincide with the theoretically expected one. Thus, we restrict ourselves
to integer computations only.

While discussing termination of integer computations the following question
should be asked: what conditions on the queries should be assumed, such that
the queries will terminate. We refer to this question as to {\em termination
inference problem}.
We illustrate this notion with the following example:
\begin{example}
\begin{eqnarray*}
&& p(X) \leftarrow X < 7, X1\;\mbox{\sl is}\;X+1, p(X1).
\end{eqnarray*}
This program terminates for queries $p(X)$, for all integer values
of $X$.  Thus, the answer for the termination inference problem is 
the condition ``{\sl true}''.
$\hfill\Box$\end{example}

This example also hints why the traditional approaches
to termination analysis fail to prove termination of this example. These
approaches are mostly based on the notion of {\em level mapping}, that
is a function from the set of all possible atoms to the natural numbers,
and should decrease while traversing the rules. In our case, such
a level mapping should depend on $X$, but $X$ can be negative as well!

Two approaches for solving this problem are possible. First, once can
change the definition of the level mapping to map atoms to integers. However,
integers are, in general, not well-founded. Thus, to prove termination
one should prove that the mapping is to some well-founded subset of integers.
In the example above $(-\infty,7)$ forms such a subset with an ordering
$\succ$, such that $x \succ y$ if $x < y$, with respect to the usual ordering
on integers.

The second approach, that we present in the paper, does not require
changing the definition of level mapping. Indeed,
the level mapping as required exists. It maps
$p(X)$ to $7-X$ if $X<7$ and to $0$ otherwise. This level mapping decreases
while traversing the rule, i.e., the size of $p(X)$, $7-X$, 
is greater than the size of $p(X1)$, $6-X$, thus, proving termination.
We present a transformation that allows to define such a level mappings
in an automatic way. The transformation presented allows to incorporate
techniques of~\cite{Dershowitz:Lindenstrauss:Sagiv:Serebrenik}, such as
level mapping inference, in the well-known framework of the acceptability
with respect to a set~\cite{DeSchreye:Verschaetse:Bruynooghe,%
Decorte:DeSchreye:98}. This integration provides not only a 
better understanding of termination behaviour of integer computations, 
but also the possibility to perform the analysis automatically
as in Decorte {\em et al.}~\cite{Decorte:DeSchreye:Vandecasteele}.

The rest of the paper is organised as following. After making some
preliminary remarks we present in Section 3 our transformation---first 
by means of an example, then more formally. In Section 4 we discuss 
more practical issues and present  the algorithm, implementing the
termination inference. In Section 5 we discuss further extensions,
such as proving termination of programs depending in numerical computations
as well as the symbolic ones.
Then we  review the related work and conclude.

\section{Preliminaries}
\subsection{Logic Programming}
We follow the standard notation for terms and atoms. A {\em query} is a 
finite sequence of atoms. Given an atom $A$, $\mbox{\sl rel}(A)$ denotes
the predicate occuring in $A$. $Term_P$ and $Atom_P$ 
denote, respectively, sets of all terms and atoms that can be constructed from
the language underlying $P$. The extended Herbrand Universe $U^E_P$ (the 
extended Herbrand base $B^E_P$) is a quotient set of $Term_P$ ($Atom_P$) modulo
the variant relation. 

We refer to an SLD-tree constructed using the left-to-right selection rule of
Prolog, as an LD-tree. We will say that a goal $G$ {\it LD-terminates} for
a program $P$, if the LD-tree for $(P,G)$ is finite.

The following definition is borrowed from~\cite{Apt:Book}.
\begin{definition}
Let $P$ be a program and $p$, $q$ be predicates occuring in it.
\begin{itemize}
\item We say that {\em $p$ refers to $q$ in $P$\/} if there is a clause in $P$
that uses $p$ in its head and $q$ in its body.
\item We say that {\em $p$ depends on $q$ in $P$\/} and write $p\sqsupseteq q$,
if $(p,q)$ is in the transitive, reflexive closure of the relation 
{\em refers to}.
\item We say that {\em $p$ and $q$  are mutually recursive\/} and write $p\simeq q$, if $p\sqsupseteq q$ and $q\sqsupseteq p$. We also write
$p\sqsupset q$ when $p\sqsupseteq q$ and $q\not\sqsupseteq p$.
\end{itemize}
\end{definition}
\subsection{Termination analysis}
In this subsection we recall some basic notions, related to termination
analysis. A {\em level mapping} is a function 
$\mid\cdot\mid: B^E_P\rightarrow {\cal N}$, where ${\cal N}$ is the set of 
the natural numbers.

The following definition generalises the notion of acceptability with 
respect to a set~\cite{DeSchreye:Verschaetse:Bruynooghe,Decorte:DeSchreye:98} 
by extending it to mutual recursion, using
 the standard notion of mutual recursion~\cite{Apt:Book}.

\begin{definition}
Let $S$ be a set of atomic queries and $P$ a definite  
program. $P$ is {\em acceptable with respect to  $S$} if there exists a 
level mapping $\mid\cdot\mid$ such that
\begin{itemize}
\item for any $A\in\mbox{\sl Call}(P,S)$ 
\item for any clause $A'\leftarrow B_1,\ldots,B_n$ in $P$, such that 
$\mbox{\rm mgu}(A,A') = \theta$ exists,
\item for any atom $B_i$, such that $\mbox{\sl rel}(B_i)\simeq \mbox{\sl rel}(A)$
\item for any computed answer substitution $\sigma$ for 
$\leftarrow (B_1, \ldots, B_{i-1})\theta$:
\[\mid A \mid\;>\;\mid B_i\theta\sigma \mid\]
\end{itemize}
\end{definition}

The following proposition characterises LD-termination in terms of 
acceptability.
\begin{theorem}(cf.\ ~\cite{DeSchreye:Verschaetse:Bruynooghe})
\label{acc:term}
Let $P$ be a program. $P$ is acceptable with respect to a set of atomic
queries $S$ if and only if $P$ is LD-terminating for all queries in $S$.
\end{theorem}

We also need to introduce the notion of interargument relations.
\begin{definition}~\cite{Decorte:DeSchreye:Vandecasteele}
Let $P$ be a definite program, $p/n$ a predicate in $P$. 
An {\em interargument relation} for $p/n$ is a relation 
$R_p \subseteq {\cal N}^n$. 
$R_p$ is a {\em valid interargument relation for $p/n$ with respect to a norm
$\|\cdot\|$} if and only if for every 
$p(t_1,\ldots,t_n)\in \mbox{\sl Atom}_P\;\mbox{\rm : if}\;\;
P\models p(t_1,\ldots,t_n)$ then $(\|t_1\|,\ldots,\|t_n\|)\in R_p$. 
\end{definition}

To characterise program transformations Bossi and Cocco~\cite{Bossi:Cocco}
introduced the following notion for a program $P$ and a query $Q$.
${\cal M}[\![P]\!](Q) =$
\begin{eqnarray*}
&&\{\sigma\mid\mbox{\rm there is a successful LD-derivation of $Q$ and $P$ with c.a.s. $\sigma$}\} \\
&& \cup \{\bot\mid\mbox{\rm there is an infinite LD-derivation of $Q$ and $P$}\} 
\end{eqnarray*}

\section{Methodology}
In this section we introduce our methodology using a simple example.
In the subsequent section we formalise it and discuss 
different extensions.

Our first example generates an oscillating sequence like $-2,4,-16,\ldots$
and stops if the generated value is greater than $1000$ or smaller than 
$-1000$. The treatment is done first on the intuitive level.
\begin{example}
\label{example:osc}
We are interested in proving termination of the set of queries 
$S = \{p(X)\mid X\;\mbox{\rm is an integer}\}$ 
with respect to the following program.
\begin{eqnarray*}
&& p(X) \leftarrow X > 1, X < 1000, X1\;\mbox{\sl is}\;-X*X, p(X1). \\
&& p(X) \leftarrow X < -1, X > -1000, X1\;\mbox{\sl is}\;X*X, p(X1). 
\end{eqnarray*}
The direct attempt to define the level mapping of $p(X)$ as $X$ fails,
since $X$ can be positive as well as negative. Thus, a more complex
level mapping should be defined. We start with some observations.

The first clause is applicable if $1 < X < 1000$, the second one,
if $-1000 < X < -1$. Observe that termination of $p(X)$ for
$X\leq -1000$, $-1 \leq X \leq 1$ or $X\geq 1000$ is trivial.
Moreover, if the first clause is applied, then
for the recursive call $p(X1)$ it holds that $-1000 < X1 < 1$. Similarly,
if the second clause applied, then for the recursive call $p(X1)$ holds
that $1 < X1 < 1000$. We use this observation and replace a predicate $p$
with two new predicates $p^{1 < X < 1000}$ and $p^{-1000 < X < -1}$,
such that $p^{1 < X < 1000}$ is called if $p(X)$ is called and $0 < X < 1000$
holds and $p^{-1000 < X < -1}$ is called if $p(X)$ is called and 
$-1000 < X < -1$ holds. The following program is obtained:
\begin{eqnarray*}
&& p^{1 < X < 1000}(X) \leftarrow X > 1, X < 1000, X1\;\mbox{\sl is}\;-X*X, p^{-1000 < X < -1}(X1). \\
&& p^{-1000 < X < -1}(X) \leftarrow X < -1, X > -1000, X1\;\mbox{\sl is}\;X*X, p^{1 < X < 1000}(X1). 
\end{eqnarray*}
Now we can define two {\em different} level mappings, one for atoms of
$p^{1 < X < 1000}$ and another one for atoms of $p^{-1000 < X < -1}$.
Let $\mid p^{1 < X < 1000}(X)\mid\;= 1000 - X$ and
let $\mid p^{-1000 < X < -1}(X)\mid\;= 1000 + X$. We verify that
the transformed program is acceptable with respect to 
$S' = \{p^{1 < X < 1000}(X) \mid 1 < X < 1000\} \cup
\{p^{-1000 < X < -1}(X)\mid -1000 < X < -1\}$ via the 
specified level mappings.
This will imply termination of the transformed program with respect to
these queries, and thus, termination of the original program with respect to 
$S$.

Indeed, consider first queries of the form $p^{1 < X < 1000}(n)$ for 
$1 < n < 1000$. The only clause that its head can be unified with this
query is the first clause and the only atom of a predicate mutually
recursive with $p^{1 < X < 1000}(X)$ is $p^{-1000 < X < -1}(m)$. Then, the
following should hold:
$\mid p^{1 < X < 1000}(n)\mid\;>\;\mid p^{-1000 < X < -1}(m)\mid$, i.e.,
$1000 - n > 1000 + m$. Recall that $n > 1$ and $m = -n^2$, thus,
$1000 - n > 1000 - n^2$ which is true for $n > 1$. Similarly, for
queries of the form $p^{-1000 < X < -1}(n)$, the acceptability
condition is reduced to $1000 + n > 1000 - n^2$ which is true for $n < -1$.
$\hfill\Box$\end{example}

The intuitive presentation above hints to the major issues to be discussed
in the following sections: how the cases as above can be extracted from the 
program, and how, given the cases extracted, the program should be transformed?
Before discussing the answers to these questions we present some basic
notions.

\subsection{Basic notions}

In this section we formally introduce some notions 
that the further analysis will be based on.
\eat{
First of all, while discussing integer computations and loops we have
to define them formally. Intuitively, a loop is a recursive part of program. 
WRONG!!!
\begin{definition}
Let $P$ be a program let $S\subseteq P$.
$S$ is called {\em loop} if there exists predicate $p$ in $S$,
such that
\begin{itemize}
\item exists $q$, $p$ refers to $q$ and $q\simeq p$
\item for every $q$, such that $p$ refers to $q$ and $q\simeq p$
and every clause $c$ with $q$ in its head, $c\in S$.
\end{itemize}
\end{definition}

\begin{definition}~\cite{Dershowitz:Lindenstrauss:Sagiv:Serebrenik}
A loop $S$ is called {\em numerical\/} if there is a clause \[H\leftarrow B_1,\ldots,
B_n\] in $S$, such that for some $i$, 
$B_i\equiv \mbox{{\sl Var\ }{\tt is\ }{\sl Exp}}$, and either {\sl Var} is 
equal to some argument of $H$ or {\sl Exp} is an arithmetic expression
involving a variable that is equal to some argument of $H$.
\end{definition}

\begin{example}
The program presented in Example~\ref{example:osc}
is a numerical loop.
$\hfill\Box$\end{example}

As shown in~\cite{Dershowitz:Lindenstrauss:Sagiv:Serebrenik}
termination of general
numerical computations may be contr-intuitive. Thus, we restrict
ourselves only to integer loops, that is numerical loops 
involving integer constants and arithmetical calculations over integers:
\begin{definition}~\cite{Dershowitz:Lindenstrauss:Sagiv:Serebrenik}
A program $P$ is {\em integer-based} if, given a query such that all 
numbers appearing in it are integers, all sub-queries that arise have this 
property as well.
\end{definition}
\begin{definition}~\cite{Dershowitz:Lindenstrauss:Sagiv:Serebrenik}
A numerical loop $S$ in a program $P$ is called an {\em integer\/} loop if $P$
is integer-based.
\end{definition}

\begin{example}
The program presented in Example~\ref{example:osc}
is an integer loop.
$\hfill\Box$\end{example}
}
Recall that the aim of our analysis, given a predicate and a query,
is to find a sufficient condition for termination of this query with respect to
this program. Thus, we need to define a notion of a termination condition.
To do so we start with a number of auxiliary definitions.

\begin{definition}
Let $p$ be a predicate of arity $n$. Then, $\$1^p,\ldots,\$n^p$ are 
called {\em argument positions denominators}.
\end{definition}
If the predicate is clear from the context the superscripts will be omitted.
\begin{definition}
Let $P$ be a program, $S$ be a set of queries. An argument position $i$
of a predicate $p$ is called {\em integer argument position}, if for 
every $p(t_1,\ldots,t_n)\in \mbox{\sl Call}(P,S)$, $t_i$ is an integer.
\end{definition}
Argument positions denominators corresponding to integer argument positions
will be called integer argument positions denominators.

An {\sl integer inequality} is an atom of one of the following forms
$\mbox{\sl Exp1} > \mbox{\sl Exp2}$,
$\mbox{\sl Exp1} < \mbox{\sl Exp2}$,$\mbox{\sl Exp1} \geq \mbox{\sl Exp2}$ or
$\mbox{\sl Exp1} \leq \mbox{\sl Exp2}$, where $\mbox{\sl Exp1}$ and
$\mbox{\sl Exp2}$ are numerical expressions, i.e., are constructed from
integers, variables and the four operations of arithmetics. A {\sl symbolic
inequality over the arguments of a predicate $p$} is constructed similarly 
to an integer inequality. However, instead of variables, integer 
argument positions denominators are used. 

\begin{example}
$X > 0$ and $Y \leq X + 5$ are integer inequalities. Given a predicate $p$
of arity 3, having only integer argument positions 
$\$1^p > 0$ and $\$2^p \leq \$1^p + \$3^p$ are symbolic 
inequalities over the arguments of $p$.
$\hfill\Box$\end{example}

Disjunctions of conjunctions based on integer inequalities are called 
{\sl integer conditions}. Similarly,  propositional calculus formulae 
based on symbolic inequalities over the arguments of a same predicate 
are called {\sl symbolic conditions over the 
integer arguments of this predicate}. 
\begin{example}
$X > 0 \wedge Y \leq X + 5$ is an integer condition. Given a predicate $p$
as above $\$1^p > 0 \wedge \$2^p \leq \$1^p + \$3^p$ is a symbolic condition
over the integer arguments of $p$.
$\hfill\Box$\end{example}

\begin{definition}
Let $p(t_1,\ldots,t_n)$ be an atom and let $c_p$ be a symbolic condition
over the arguments of $p$. An {\em instance of the condition with respect to an atom},
denoted $c_p(p(t_1,\ldots,t_n))$,
is obtained by replacing the 
argument positions denominators with the corresponding
arguments, i.e., $\$i^p$ with $t_i$.
\end{definition}

\begin{example}
Let $p(X, Y, 5)$ be an atom and let $c_p$ be 
$(\$1^p > 0) \wedge (\$2^p \leq \$1^p + \$3^p)$. Then, $c_p(p(X, Y, 5))$
is the integer conjunction $(X > 0) \wedge (Y \leq X + 5)$.
$\hfill\Box$\end{example}

Now we are ready to define {\em termination condition\/} formally.

\begin{definition}
Let $P$ be a program, and $Q$ be a query.
A symbolic condition $c_{\mbox{\sl rel}(Q)}$ is a 
{\em termination condition} for $Q$ if given that $c_{\mbox{\sl rel}(Q)}(Q)$
holds, $Q$ left-terminates with respect to $P$.
\end{definition}

A termination condition for Example~\ref{example:osc} is {\sl true}, i.e.,
the query $\leftarrow p(X)$, terminates with respect to the program
for all integer $X$. Obviously this is not always the case.
\begin{example}
Consider the following program.
\label{example:simple}
\begin{eqnarray*}
&& q(X) \leftarrow X > 0, X \leq 5, q(X).\\
&& q(X) \leftarrow X > -5.
\end{eqnarray*}
This program terminates for queries $\leftarrow p(n)$, such that $n \leq -5$,
since no rule is applicable, or $-5 < n \leq 0 \vee n > 5$, since 
repeated rule application
in this case is finite. Thus, a termination condition for the goal
$\leftarrow q(X)$ is $\$1\leq 0 \vee \$1 > 5$.
$\hfill\Box$\end{example}

\subsection{Types}
\label{subsection:types}
In this subsection we discuss inferring what values  integer
arguments can take
during traversal of the rules, i.e., the ``case analysis'' performed
in Example~\ref{example:osc}. This information is crucial for
defining level-mappings. 
Example~\ref{example:osc} provides already the underlying 
intuition---``cases'' are types, i.e., calls of the predicate
$p^c$ are identical to the calls of the predicate $p$, where $c$ holds
for its arguments. More formally we define a notion 
of {\em set of adornments},
later  we specify when it is {\em guard-tuned} and we show how such a guard-tuned
set of adornments can be constructed.

\begin{definition}
\label{def:set:of:adornments}
Let $p$ be a predicate, and let $c_1,\ldots,c_n$ be symbolic conditions
over the integer arguments of $p$. The set ${\cal A}_p = \{c_1,\ldots,c_n\}$
is called {\em set of adornments for $p$} if for all $i,j$ such that 
$1\leq i < j\leq n$, $c_i \wedge c_j = \mbox{\sl false}$ and
$\bigvee_{i=1}^n c_i = \mbox{\sl true}$.
\end{definition}

A set of adornments partitions the domain for (some of) the integer variables
of the predicate.

\begin{example}
\label{example:osc:cases}
Let $P$ be as in Example~\ref{example:osc}. The following
are examples of sets of adornments: $\{\$1\leq 100,\$1 > 100\}$ and
$\{(\$1\leq -1000) \vee (-1 \leq \$1 \leq 1) \vee (\$1\geq 1000),
-1000 < \$1 < -1,1 < \$1 < 1000\}$.
$\hfill\Box$\end{example}

\subsection{Program transformation}
The next question that should be answered is how the program
should be transformed given a set of adornments. After this transformation
$p^c(X_1,\ldots,X_n)$ will behave with respect to the transformed program
exactly as $p(X_1,\ldots,X_n)$ does, for all calls that satisfy the condition
$c$. Intuitively, we replace each call to the predicate $p$ 
in the original program by a number of possible calls in the transformed one.
To define a transformation formally we introduce the following definition:

Let $H\leftarrow B_1,\ldots,B_n$ be a rule. $B_1,\ldots,B_i$, $1\leq i\leq n$,
is called {\em prefix of the rule}, if for all $j$, $1\leq j\leq i$,
$B_j$ is an integer inequality and the only variables appearing in
its arguments are variables of $H$. $B_1,\ldots,B_i$
is called {\sl the maximal prefix} of the rule,  if it is a prefix and 
$B_1,\ldots,B_i,B_{i+1}$ is not a prefix.

Observe, that since a prefix constrains only variables appearing in the head
of a clause there exists a symbolic condition over the
arguments of the predicate of the head, such that the prefix is its instance
with respect to the head. Note, that in general, this symbolic condition is 
not necessarily unique.
\begin{example}
Consider the following program: $p(X, Y, Y) \leftarrow Y > 5.$
The only prefix of this rule is $Y > 5$. There are two symbolic
conditions over the arguments of $p$, $\$2 > 5$ and $\$3 > 5$,
such that $Y > 5$ is their instance
with respect to $p(X, Y, Y)$. 
$\hfill\Box$\end{example}

The following notion, borrowed from
\cite{Dershowitz:Lindenstrauss:Sagiv:Serebrenik}, 
guarantees uniqueness of such symbolic conditions. In this case we 
say that the symbolic condition {\em corresponds\/} to the prefix.

\begin{definition}~\cite{Dershowitz:Lindenstrauss:Sagiv:Serebrenik}
A rule $H\leftarrow B_1,\ldots,B_n$ is called {\em partially normalised}
if all integer argument positions in $H$ are occupied by distinct
variables\footnote{If such a rule has only integer arguments Apt 
{\em et al.}~\cite{Apt:Marchiori:Palamidessi} call it {\em homogeneous}.}.
\end{definition}

We will also say that a program $P$ is partially normalised
if all the rules in $P$ are partially normalised.
After integer argument positions are identified a
program can be easily rewritten to partially normalised form. 

Now we are ready to present the transformation formally.

\begin{definition}
Let $P$ be a program and let $p$ be a predicate in it. Let 
${\cal A} = \bigcup_{q\in P} {\cal A}_q$
be a set of possible adornments for $P$. Then, the program $P^a$,
called {\em adorned with respect to $p$}, is obtained by two steps as following:
\begin{enumerate}
\item For every rule $r$ in $P$ and for every subgoal $q(t_1,\ldots,t_n)$, $p\simeq q$, in $r$
\begin{itemize}
\item[] For every $A\in {\cal A}_q$ 
\item[] Replace $q(t_1,\ldots,t_n)$ by $q^A(t_1,\ldots,t_n)$.
\end{itemize}
\item For every newly obtained rule $r$ 
\begin{itemize}
\item[] Are adornments and inequalities in the body of $r$ consistent? $\hfill *$
\item[] \begin{itemize}
\item[] If not---reject the rule.
\end{itemize}
\item[] If $r$ defines some $q$, $q\simeq p$
\item[] \begin{itemize}
\item[] Get as adornments of the head of $r$ all $A\in {\cal A}_{q}$,
that are consistent with comparisons of the maximal prefix of $r$ and
adornments of the body of $r$.
\end{itemize}\end{itemize}
\end{enumerate}
\end{definition}

\begin{example}
\label{example:osc:adornments}
Continue Example~\ref{example:osc}. The sets of adornments presented
in Example~\ref{example:osc:cases} are used. 
With the first set of adornments in Example~\ref{example:osc:cases} we obtain
the program:
\begin{eqnarray*}
&& p^{\$1\leq 100}(X) \leftarrow X > 1, X < 1000, X1\;\mbox{\sl is}\;-X*X, p^{\$1\leq 100}(X1). \\
&& p^{\$1 > 100}(X) \leftarrow X > 1, X < 1000, X1\;\mbox{\sl is}\;-X*X, p^{\$1\leq 100}(X1). \\
&& p^{\$1\leq 100}(X) \leftarrow X < -1, X > -1000, X1\;\mbox{\sl is}\;X*X, p^{\$1\leq 100}(X1). \\ 
&& p^{\$1\leq 100}(X) \leftarrow X < -1, X > -1000, X1\;\mbox{\sl is}\;X*X, p^{\$1 > 100}(X1). 
\end{eqnarray*}
If the second set of adornments is used, the following program is obtained:
\begin{eqnarray*}
\hspace{1.7cm}&& p^{1 < \$1 < 1000}(X) \leftarrow X > 1, X < 1000, X1\;\mbox{\sl is}\;-X*X, \\
&& \hspace{2.8cm} p^{-1000 < \$1 < -1}(X1). \\
&& p^{1 < \$1 < 1000}(X) \leftarrow X > 1, X < 1000, X1\;\mbox{\sl is}\;-X*X,\\
&& \hspace{2.8cm} p^{(\$1\leq -1000) \vee (-1 \leq \$1 \leq 1) \vee (\$1\geq 1000)}(X1). \\
&& p^{-1000 < \$1 < -1}(X) \leftarrow X < -1, X > -1000, X1\;\mbox{\sl is}\;X*X,\\
&& \hspace{3.2cm} p^{1 < \$1 < 1000}(X1). \\
&& p^{-1000 < \$1 < -1}(X) \leftarrow X < -1, X > -1000, X1\;\mbox{\sl is}\;X*X, \\
&& \hspace{3.2cm}p^{(\$1\leq -1000) \vee (-1 \leq \$1 \leq 1) \vee (\$1\geq 1000)}(X1).
\end{eqnarray*}
$\hfill\Box$\end{example}
Correctness of the transformation should be proved. First of all,
finiteness of the number of clauses, the number of subgoals in a clause and
the number of elements in an adornment 
ensures that the transformation always terminates.
Second, we need to prove that the transformation preserves termination.

Adorning clauses introduces new predicates. This means that the query
$Q$ gives rise to a number of different queries. Clearly, termination
of all of these queries with respect to $P^{a}$ is equivalent to termination of
$Q$ with respect to $P^{a}$ augmented by a set of the clauses, such that
for every $p\simeq \mbox{\sl rel}(Q)$ and for every $A\in {\cal A}_p$
the clause $p(X_1,\ldots,X_n) \leftarrow p^A(X_1,\ldots,X_n)$ is added. 
We call this extended program $P^{ag}$.

\begin{lemma}
Let $P$ be a program, and let $Q$ be a query. Let $P^{ag}$ be a program 
obtained as described above.
Then, ${\cal M}[\![P^{ag}]\!](Q) \subseteq {\cal M}[\![P]\!](Q)$.
\end{lemma}
\begin{proof}
Proof is done similarly to~\cite{Lindenstrauss:Sagiv:Serebrenik:L}.
Replace each call to adorned predicate $p^A$ in the body of clauses originating
from $P^{a}$ by the corresponding call to $p$. Call the obtained program
$P^{agw}$. Since $P^{ag}\setminus P^a$ has a clause for every adornment
in ${\cal A}_p$, every path in the LD-tree of $Q$ w.r.t. $P^{ag}$ has
a corresponding path in the LD-tree of $Q$ w.r.t. $P^{agw}$. Thus, 
${\cal M}[\![P^{ag}]\!](Q) \subseteq {\cal M}[\![P^{agw}]\!](Q)$.

Similarly, every path in the LD-tree of $Q$ w.r.t. $P^{agw}$ has
a corresponding path in the LD-tree of $Q$ w.r.t. $P$. Indeed, every call
to $p$ on the path of the LD-tree of $Q$ w.r.t. $P^{agw}$ is followed by
calls to all adorned versions of $p$ via rules originating from
$P^{ag}\setminus P^{a}$, and subsequently to the rules of those predicates.
However, in $P$ these rules are directly defining $P$. Thus, 
${\cal M}[\![P^{agw}]\!](Q) \subseteq {\cal M}[\![P]\!](Q)$.

We conclude, that ${\cal M}[\![P^{ag}]\!](Q) \subseteq {\cal M}[\![P]\!](Q)$.
$\hfill\blacksquare$\end{proof}

The second direction of the containment depends on
the consistency check strategy applied at the point marked by $*$
in the definition of $P^a$.
\begin{example}
Let $Q$ be $p(X)$ and let $P$ be the following program
\[p(X)\leftarrow X > 0, q(X), X < 0.\;\;\;\;\;\;
q(X)\leftarrow X > 0, p(X).\]
Predicates $p$ and $q$ are mutually recursive. Thus, both of them should
be adorned. Let ${\cal A}_p$ be $\{\$1 > 0, \$1 \leq 0\}$ and
${\cal A}_q$ be ${\cal A}_q = \{\$1 > 0, \$1 \leq 0\}$.
The following program  is obtained after the first step of the adorning process.
\begin{eqnarray*}
&& p(X)\leftarrow X > 0, q^{\$1 > 0}(X), X < 0.\;\;\;\;\;\;q(X)\leftarrow X > 0, p^{\$1 > 0}(X).\\
&& p(X)\leftarrow X > 0, q^{\$1 \leq 0}(X), X < 0.\;\;\;\;\;\;q(X)\leftarrow X > 0, p^{\$1 \leq 0}(X).
\end{eqnarray*}
The second step of the adorning process should infer adornments for the
heads of the clauses, possibly rejecting the inconsistent ones.
If the inference technique is eager, i.e., tries to use
all the information it has in the body constraints and adornments
of body subgoals, a program consisting only of one rule, namely
$q^{\$1 > 0}(X)\leftarrow X > 0, p^{\$1 > 0}(X)$, is obtained.
Other clauses are rejected because inconsistency of the
set of built-in comparisons and adornments applied to the corresponding atoms
is discovered. Thus, the extended program is the following one:
\begin{eqnarray*}
&& \;\;\;\;\;\;q^{\$1 > 0}(X)\leftarrow X > 0, p^{\$1 > 0}(X).\\
&& q(X) \leftarrow q^{\$1 > 0}(X). \;\;\;\;\;\;p(X) \leftarrow p^{\$1 > 0}(X).\\
&& q(X) \leftarrow q^{\$1 \leq 0}(X). \;\;\;\;\;\;p(X) \leftarrow p^{\$1 \leq 0}(X). 
\end{eqnarray*}
The query $p(X)$ terminates with respect to this program, while it does not
terminate with respect to the original one. This example shows that eager 
inference technique can actually improve termination.

In order termination to be preserved a weaker inference engine should be used.
For example, one can use an inference technique that considers inequalities
only of the maximal prefix. If this technique is used, the following
program is obtained after extension:
\begin{eqnarray*}
&& \;\;\;\;\;\;p^{\$1 > 0}(X)\leftarrow X > 0, q^{\$1 > 0}(X), X < 0. \\
&& \;\;\;\;\;\;q^{\$1 > 0}(X)\leftarrow X > 0, p^{\$1 > 0}(X).\\
&& q(X) \leftarrow q^{\$1 > 0}(X). \;\;\;\;\;\; p(X) \leftarrow p^{\$1 > 0}(X). \\
&& q(X) \leftarrow q^{\$1 \leq 0}(X). \;\;\;\;\;\;p(X) \leftarrow p^{\$1 \leq 0}(X).
\end{eqnarray*}
The query $p(X)$ does not terminate with respect to this program, just as it
does not terminate with respect to the original one. The following lemma shows
that if this weaker inference technique is used, termination is preserved.
$\hfill\Box$\end{example}

Observe that it is known that unfolding preserves computed answer 
substitutions~\cite{Bossi:Cocco:Basic,Bossi:Cocco}, thus in order to
prove the second direction of the containment we have to prove that 
termination is preserved. Unlike the previous lemma that can be 
established for an arbitrary set of symbolic conditions, used
as adornments, this lemma holds 
only sets of adornments as defined in Definition~\ref{def:set:of:adornments}.

\begin{lemma}
Let $P$ be a program, and let $Q$ be a query. Let $P^{ag}$ be a program 
obtained as described above with respect to a set of adornments and consistency
checking with respect to maximal prefixes.
Then, ${\cal M}[\![P]\!](Q) \subseteq {\cal M}[\![P^{ag}]\!](Q)$.
\end{lemma}
\begin{proof}
Assume that $Q$ terminates w.r.t. $P^{ag}$ and does not terminate w.r.t. $P$.
Let $H\leftarrow B_1,\ldots,B_n$ be a clause in $P$, such that $Q$ can be 
unified with $H$ and the application of this rule starts an infinite
branch in the LD-tree.

In the resolution w.r.t. $P^{ag}$ the only clauses to be applied to
resolve with $Q$ are those of $P^{ag}\setminus P^{a}$. This will reduce
the query to queries of the form $q^A(t_1, \ldots, t_k)$, where $q^A$
are adorned versions of the predicate $q$ of $Q$ and $t_1,\ldots,t_k$ are
arguments of $Q$.

First of all, we prove that for some adornment of $q$ there exists an
adorned variant of 
$H\leftarrow B_1,\ldots,B_n$ exists in $P^{ag}$. For the sake of contradiction
assume that this does not hold, i.e., there is no 
possible adornment of atoms of predicates mutually
recursive with $\mbox{\sl rel}(H)$ that is consistent with comparisons
of the maximal prefix of the clause and with one of the possible adornments
for $\mbox{\sl rel}(H)$.
This verbal description above can be rewritten in the following way:
\[
\begin{array}{lll}
\mbox{\sl false}&=& A_1 \& C \& B_{11} \& \ldots \& B_{n1} \\
&& \ldots \\
\mbox{\sl false}&=& A_n \& C \& B_{1m_1} \& \ldots \& B_{nm_n} 
\end{array}
\]
where $A_1,\ldots,A_n$ are all possible adornments for $\mbox{\sl rel}(H)$,
$C$ is a conjunction of comparisons of the maximal prefix of the clause and 
$B_{11},\ldots,B_{nm_n}$ are all possible adornments for the atoms of 
predicates mutually recursive with $\mbox{\sl rel}(H)$ and appearing in the 
body of the clause. Disjunction of these conjunctions is, on the one
hand, $\mbox{\sl false}$ and on the other hand, $C$ (since
${\cal A}_p$ is complete). Thus, $C = \mbox{\sl false}$.

Observe, that comparisons of the maximal prefix are not affected
by the transformation. Thus, the body of the clause above starts with
a sequence of inconsistent comparisons. Since comparisons are part of the
maximal prefix, inconsistency will be discovered before any atom
other than a comparison is reached. Thus, any application of this clause
will cause a failure and it cannot start an infinite branch of the LD-tree.
Thus, the adorned version of $H\leftarrow B_1,\ldots,B_n$ exists in $P^{ag}$.
Let $H'\leftarrow B'_1,\ldots, B'_n$ be this adorned version, where $B'_i$
denotes either an adorned version of $B_i$, if $B_i$ was adorned,
and is identical to $B_i$ otherwise.

Let $B'_j$ be the first (while going from left to right) adorned atom in
the clause body. Since transformation does not affect the preceding body atoms
the corresponding queries are identical w.r.t. $P$ and w.r.t. $P^{ag}$.
Let $Q'_j$ be a query corresponding to $B'_j$, and let $Q_j$ be a query
corresponding to $Q_j$. We have to prove that $Q_j$ terminates.

Assume that $Q_j$ does not terminate. Let $G\leftarrow \ldots$ be a clause 
in $P$, such that $Q_j$ is resolved with on the infinite branch of the LD-tree.
By the previous claim there is an adorned version of this clause that belongs 
to $P^{ag}$. If there is
no adorned version of this clause with the adornment of $Q'_j$ then
by reasoning similar to above one can conclude that this adornment is
inconsistent with the comparisons of the maximal prefix. Since those
are not affected by the transformation and are identical in the clause of
$P$ and in the corresponding clause of $P^{ag}$.
Thus, any application of this clause will cause a failure and it cannot 
continue an infinite branch of the LD-tree. This means, that there
exists an adorned variant of the clause, such that its head can be unified
with $Q'_j$. Since computed answer substitutions of queries with respect to
$P$ and of $P^{ag}$
are identical (follows from the fact that unfolding preserves computed
answer substitutions~\cite{Kawamura:Kanamori}) the same reasoning 
can be done for any of the subsequent calls and clauses, i.e., we will mimic
the resolution started by $Q_j$ w.r.t. $P$ by a resolution that is started
by $Q'_j$ w.r.t. $P^{ag}$. Since any resolution of $Q'_j$ w.r.t. $P^{ag}$
is finite, contradiction to the assumption is obtained.

Since computed answers are preserved the same claim can be proved also for
the queries, originating from other atoms that $B_j$, thus, completing the 
proof.
$\hfill\blacksquare$\end{proof}

The following theorem summarises 
lemmas above, for the case of maximal prefixes.
\begin{theorem}
\label{theorem:pres}
Let $P$ be a program, let $Q$ be a query and let ${\cal A}$ be a 
set of adornments. Let $P^{ag}$ be a program 
obtained as described above with respect to ${\cal A}$.
Then, ${\cal M}[\![P]\!](Q) = {\cal M}[\![P^{ag}]\!](Q)$.
\end{theorem}

This theorem has two corollaries. The first one 
establishes a relation between the termination condition and 
the adornments and the second one allows to reason
on the termination with respect to the original program $P$.

\begin{corollary}
\label{cor:nonrec}
Let $P$ be a program, let $Q$ be a query and let ${\cal A}$ be a 
set of adornments. Let 
\[A = \{a\mid a\in {\cal A}, \mbox{\sl for all}\;q\;\mbox{\sl rel}(Q)^a \sqsupseteq q,\;q\;\mbox{\sl is not recursive in
 $P^a$}\}.\] Then $\bigvee_{a\in A}$ is a termination condition for $P$.
\end{corollary}

\begin{example}
Continue Example~\ref{example:osc:adornments}. In the program
obtained with respect to the second set of adornments, predicate
$p^{(\$1\leq -1000) \vee (-1 \leq \$1 \leq 1) \vee (\$1\geq 1000)}$
satisfies the Corollary.
Thus, $(\$1\leq -1000) \vee (-1 \leq \$1 \leq 1) \vee (\$1\geq 1000)$
is a termination condition for $p(X)$ with respect to the program
presented in Example~\ref{example:osc}.
$\hfill\Box$\end{example}

Theorem~\ref{theorem:pres} implies that a program $P$ is  
LD-terminating with respect to all queries in a set of atomic queries $S$ if
and only if $P^{ag}$, constructed as above, is acceptable with respect to $S$.
However, the later one is equivalent to acceptability of $P^{a}$ with
respect to $\{q^A(t_1,\ldots,t_n)\mid q(t_1,\ldots,t_n)\in S, A\in {\cal A}_q\}$.

\begin{corollary}
Let $P$ be a program, let $S$ be a set of atomic queries and let 
${\cal A} = \bigcup_{Q\in (S), q\simeq \mbox{\sl rel}(Q)}{\cal A}_q$
be a set of adornments. Let $P^{a}$ be
obtained with respect to ${\cal A}$.
Then, $P$ is LD-terminating with respect to all queries in $S$ if
and only if $P^{a}$ is acceptable with respect to $\{q^A(t_1,\ldots,t_n)\mid p(t_1,\ldots,t_n)\in S, A\in {\cal A}_q\}$.
\end{corollary}

This corollary allows to complete the termination proof for 
Example~\ref{example:osc}.

\begin{example}
\label{example:osc:lm}
The transformed program $P^{a}$ (with respect to $\{-1000 < \$1 < -1,
1 < \$1 < 1000, (\$1\leq -1000) \vee (-1 \leq \$1 \leq 1) \vee (\$1\geq 1000)\}$) is presented in Example~\ref{example:osc:adornments}. We 
prove acceptability of $P^a$ with respect to the set $S = 
\{p^{1 < \$1 < 1000}(X), p^{-1000 < \$1 < -1}(X), \\p^{(\$1\leq -1000) \vee (-1 \leq \$1 \leq 1) \vee (\$1\geq 1000)}(X)\}$. Then, 
$S = \mbox{\sl Call}(P^a,S)$. Let $\mid \cdot \mid$ be the level mapping, 
defined as follows:
\begin{eqnarray*}
&& \mid p^{-1000 < \$1 < -1}(X) \mid \;\;= 
\left\{
\begin{array}{ll}
1000 + X & \mbox{\rm if}\;\;-1000 < X < -1\\
0        & \mbox{\rm otherwise}
\end{array}
\right. \\
&& \mid p^{1 < \$1 < 1000}(X) \mid\;\; = \left\{
\begin{array}{ll}
1000 - X & \mbox{\rm if}\;\;1 < X < 1000\\
0        & \mbox{\rm otherwise}
\end{array}
\right. \\
&& \mid p^{(\$1\leq -1000) \vee (-1 \leq \$1 \leq 1) \vee (\$1\geq 1000)}(X) \mid\;\; = 0
\end{eqnarray*}
We are not going to prove completely that $P^a$ is acceptable with respect
to $S$ via $\mid \cdot \mid$, but restrict ourselves
only for one call, $p^{1 < \$1 < 1000}(X)$. There are two clauses---the
first and the second one---such that their heads can be unified with
$p^{1 < \$1 < 1000}(X)$. The second clause is not recursive and the condition
holds vacuously. The first clause is recursive and the acceptability
requires $1000 - X > 1000 + X1$, where $X > 1, X < 1000$ and $X1 = - X^2$.
Substituting the last equality and simplifying one gets $X^2 > X$, that 
is true for $X > 1$. Other calls are solved similarly.
$\hfill\Box$\end{example}

\section{Practical issues}
In the previous section we have shown the transformation that allows
reasoning on termination of the numerical computations in the framework
of acceptability with respect to the set. In this section we discuss
a number of practical issues to be considered for an automated termination
analysis.

\subsection{Guard-tuned sets of adornments}
In Example~\ref{example:osc:cases} we have seen two different sets of 
adornments. Both of them are valid according to Definition~\ref{def:set:of:adornments}. However, $\{-1000 < \$1 < -1,
1 < \$1 < 1000, (\$1\leq -1000) \vee (-1 \leq \$1 \leq 1) \vee (\$1\geq 1000)\}$ is in some sense preferable to $\{\$1\leq 100,\$1 > 100\}$. There is a
number of reasons to prefer the first set to the second one. First of all,
it has {\em a declarative reading}: the sets that are constructed are
related to the constraints in the bodies of the clauses and in fact 
express conditions that, when satisfied, allow to traverse the rule. Second,
comparing the two adorned programs in Example~\ref{example:osc:adornments}
one might observe that the second program has two mutually
recursive predicates, connected by two clauses, while the first program
has not only this connection, but also self-loop on one of the predicates.

Intuitively, a set of adornments of a predicate $p$ is {\em guard-tuned} if for 
every adornment $a$ in it and every clause $c$ of the program defining $p$
the conjunction of the maximal prefix of $c$ 
and $a$ 
is either {\sl false} or the conjunction 
is identical to $a$. We will see that the set of
adornments we preferred in the discussion above is guard-tuned, while the
second set is not. 

\begin{definition}
Let $P$ be a partially normalised program, 
let $p$ be a predicate in $P$, and let ${\cal A}_p$
be a set of adornments for $p$. We say that ${\cal A}_p$
is {\em guard-tuned\/}
if for every $A\in {\cal A}_p$ and
for every rule $r\in P$ with the symbolic condition $c$
corresponding to the maximal prefix of $r$ holds
that either $c \wedge A = \mbox{\sl false}$
or $c \wedge A = A$.
\end{definition}

\begin{example}
\label{example:osc:cases2}
Consider 
the sets of adornments presented in 
Example~\ref{example:osc:cases}.
The first set of adornments is not guard-tuned while
the second one is guard-tuned.
$\hfill\Box$\end{example}

\subsection{How to construct a guard-tuned set of adornments?}
\label{subsection:inferring}
In this subsection we present a technique allowing to construct 
a guard-tuned set of adornments for a predicate $p$ given a program $P$. To do so, 
recall once more Examples~\ref{example:osc:cases} and \ref{example:osc:cases2}.
They suggest two ways of constructing such a set. The first one is: 
given a program $P$ collect the symbolic 
conditions, corresponding to the maximal prefixes of the rules defining $p$ 
(we denote this set ${\cal C}_p$) and
add the completion of the constructed disjunction. Unfortunately, this set of 
conditions is not necessarily a set of adornments and if so, it
is not necessary guard-tuned.

\begin{example}
\label{example:intersections:needed}
Consider the following program.
\begin{eqnarray*}
&& r(X) \leftarrow X > 5.\\
&& r(X) \leftarrow X > 10, r(X).
\end{eqnarray*}
Two sets of symbolic conditions can be constructed in the way described above:
$\{r^{\$1\leq 5}, r^{\$1 > 5}, r^{\$1 > 10}\}$ which is not a set of adornments
and $\{r^{\$1\leq 5}, \\ r^{\$1 > 5}\}$ which is not guard-tuned due to the second
rule of the program.
$\hfill\Box$\end{example}

Thus, we are going to use a different approach.
Once more, we start by finding ${\cal C}_p$. Let ${\cal C}_p = 
c_1, \ldots, c_n$, then let ${\cal A}_p$ be the set of 
conjunctions of $c_i$'s and their negations. We claim that the constructed 
set is
always a guard-tuned set of adornments. Before stating this formally, consider
once more Example~\ref{example:intersections:needed}.

\begin{example}
In this case, the symbolic conditions corresponding to the maximal prefixes
of the rules are $\$1 > 5$ and $\$1 > 10$. Thus, the adornments are:
$\$1 > 5 \wedge \$1 > 10, \$1 \leq 5 \wedge \$1 > 10, \$1 > 5 \wedge \$1 \leq 10, \$1 \leq 5 \wedge \$1 \leq 10$. After simplifying and 
removing the inconsistent
conjuncts:
$\$1 > 10, \$1 > 5 \wedge \$1 \leq 10, \$1 \leq 5$.
$\hfill\Box$\end{example}

\begin{lemma}
Let $P$ be a program, $p$ be a predicate in $P$ and ${\cal A}_p$ be 
constructed as described. Then ${\cal A}_p$ is a guard-tuned set of adornments.
\end{lemma}
\begin{proof}
The proof is immediate by checking the definitions
$\hfill\blacksquare$\end{proof}

\subsection{How to define a level mapping?}
One of the questions that should be answered is how the level mappings
should be generated automatically. 
The problem with defining level mappings is that they should reflect changes
on possibly negative integer arguments, on the one hand, and remain 
non-negative, on the other. We also like to remain in the
framework of level mappings on atoms defined
as linear combinations of sizes of their arguments.

We are going to solve this problem by defining different level mappings
for different adorned versions of the predicate.
The major observation underlying
the technique presented in this subsection is that if $\$1 > \$2$
appears in the adornment of a recursive clause, 
then for each call to this adorned predicate
$\$1 - \$2$ will be positive, and thus, can be used for defining a 
level mapping. More formally:
\begin{definition}
Let $p^{E_1\;\rho\;E_2}$ be an adorned predicate, where
$E_1$ and $E_2$ are expressions and $\rho\in\{>,\geq\}$. 
The {\em primitive level mapping}
is defined as:
\[\left\{
\begin{array}{ll}
(E_1 - E_2)(t_1,\ldots,t_n) & \mbox{\rm if}\;\;E_1(t_1,\ldots,t_n)\;\rho\;E_2(t_1,\ldots,t_n) \\
0         & \mbox{\rm otherwise}
\end{array}
\right. \]
\end{definition}
In most of the examples more than one conjunct will appear in the 
adornment. In this case the level mapping is defined as a linear 
combination of primitive level mappings corresponding to the conjuncts. 
Some of the conjuncts
may actually be 
disjunctions---they are ignored, since disjunctions can be introduced
only by the fact the some rule {\em cannot} be applied. 
\begin{definition}
Let $p^c$ be an adorned predicate, 
The {\em natural level mapping}
is defined as: \[\mid p^c(t_1,\ldots,t_n) \mid\;\; =
\sum_{E_1\;\rho\;E_2 \in c} c_{E_1\;\rho\;E_2} \mid p^{E_1\;\rho\;E_2}(t_1,\ldots,t_n) \mid^{\mbox{\sl pr}},\] where $c$'s are natural coefficients, $E_1$ and $E_2$ are expressions and
$\rho\in\{>,\geq\}$.
\end{definition}

\begin{example}
The level mappings used in Example~\ref{example:osc:lm} are natural
level mappings with the following coefficients:
$c_{\$1 > 1} = 0$, $c_{\$1 < 1000} = 1$,
$c_{\$1 < -1} = 0$, $c_{\$1 > -1000} = 1$. For $p^{(\$1\leq -1000) \vee (-1 \leq \$1 \leq 1) \vee (\$1\geq 1000)}$ the definition holds trivially.
$\hfill\Box$\end{example}

Technique developed by Decorte {\sl et al.}~\cite{Decorte:DeSchreye:Vandecasteele} allows to define symbolic counterparts of the level
mappings and to infer the actual values of the coefficients by solving
a system of constraints.

\subsection{Putting it all together and inferring termination constraints}
In this section we show how the steps studied so far can be combined to 
an algorithm that allows to infer termination conditions. Intuitively,
one starts with a termination condition initialised to be {\sl true}, i.e.,
assuming that 
query $Q$ terminates with respect to a program $P$ for all possible values of integer
arguments. Constraints are constructed similarly to~\cite{Decorte:DeSchreye:Vandecasteele}. 
If these can be satisfied without imposing
any additional constraints on the integer variables,
stop and report termination for 
the condition constructed so far. If these impose some constraints involving
a new integer variable, repeat the process. If neither of the cases hold,
stop and report possibility of non-termination.

Any other technique proving termination and
being able not only to claim ``termination can be proved'' or
``termination cannot be proved'' but also providing in the latter case
some constraint that, if satisfied, implies termination can be 
used instead of~\cite{Decorte:DeSchreye:Vandecasteele}.
The algorithm is presented in Figure~\ref{algo}.
\begin{figure*}[htb]
\begin{center}
\fbox{
\parbox{4.6in}{
Let $P$ be a partially normalised 
program, let $Q$ be a query and let $q$ be $\mbox{\sl rel}(Q)$.
\begin{enumerate}
\item Initialise the termination condition $c$ to be {\sl true}.
\item For each $p\simeq q$ construct ${\cal A}_p$.
\item Adorn $P$ w.r.t. $q$ and $\bigcup_{p\simeq q}{\cal A}_p$.
\item Remove ``irrelevant clauses''
\item[] \begin{itemize}
\item[] Let $A_1,\ldots,A_n \in {\cal A}_q$ be the only adornments of $q$ \\
that are consistent with $c$.
\item[] For every rule $r$ in $P^a$
\item[] \begin{itemize}
\item[] If for all $i$, $q^{A_i}\not\sqsupseteq \mbox{\sl rel}(\mbox{\sl Head}(r))$
\item[] \begin{itemize}
\item[] Remove $r$ from $P^a$
\end{itemize}
\end{itemize}
\end{itemize}
\item Define a symbolic counterparts of norms, \\ level mappings and interargument relations. 
\item Construct constraints on the symbolic variables. Obtain $S$. 
\item Solve $S$. 
\begin{enumerate}
\item Solution of $S$ doesn't produce extra constraints on variables.
\begin{itemize}
\item[] Report termination for the condition constructed so far.
\end{itemize}
\item Solution of $S$ produces extra constraints involving new
integer variables.
\begin{itemize}
\item[] Add these constraints to termination condition.
\item[] Go back to step 2.
\end{itemize}
\item Otherwise report possibility of non-termination.
\end{enumerate}
\end{enumerate}
}}
\caption{Termination Inference Algorithm}
\label{algo}
\end{center}
\end{figure*}
\vspace{-0.2in}
\begin{example}
\label{example:gcd:simplified:2}
Consider the following program.
\begin{eqnarray*}
&& q(X,Y) \leftarrow X>Y, Z\;\mbox{\sl is}\;X-Y, Y1\;\mbox{\sl is}\;Y+1, q(Z,Y1).
\end{eqnarray*}
We are interested in finding values of $X$ and $Y$ such that $q(X,Y)$
terminates. The first step of our algorithm
is inferring the sets of adornments. 
There is only one inequality in the clause, i.e., $X>Y$.
The corresponding symbolic constraint is $\$1 > \$2$. 
Thus, the inferred adornment is $\{\$1>\$2, \$1\leq\$2\}$.

The adorned version of this program is 
\begin{eqnarray*}
&& q^{\$1>\$2}(X,Y) \leftarrow X>Y, Z\;\mbox{\sl is}\;X-Y, Y1\;\mbox{\sl is}\;Y+1, q^{\$1>\$2}(Z,Y1).\\
&& q^{\$1>\$2}(X,Y) \leftarrow X>Y, Z\;\mbox{\sl is}\;X-Y, Y1\;\mbox{\sl is}\;Y+1, q^{\$1\leq\$2}(Z,Y1).
\end{eqnarray*}

There is no clause defining $q^{\$1\leq\$2}$. By Corollary~\ref{cor:nonrec}, $\$1\leq \$2$ is a termination condition. 
The recursive clause is analysed further. The level mapping is
\[\mid q^{\$1>\$2}(X,Y) \mid\;= c_{\$1>\$2} * \left\{ \begin{array}{ll}
X-Y & \mbox{\rm if}\;\;X>Y \\
0 & \mbox{\rm otherwise}\end{array} \right.\]

The 
acceptability decrease implies (see~\cite{Decorte:DeSchreye:Vandecasteele}):
\[c_{\$1>\$2}(X-Y) >c_{\$1>\$2}(X-Y) - c_{\$1>\$2}Y,\]
that is $c_{\$1>\$2} Y > 0$. Since $c_{\$1>\$2} \geq 0$, $Y>0$ should hold. 
Now we restart the whole process with respect to $Y>0$. 
The following adorned program is obtained:
\begin{eqnarray*}
&& q^{\$1>\$2,\$2>0}(X,Y) \leftarrow X>Y, Z\;\mbox{\sl is}\;X-Y, Y1\;\mbox{\sl is}\;Y+1, q^{\$1>\$2,\$2>0}(Z,Y1)\\
&& q^{\$1>\$2,\$2\leq 0}(X,Y) \leftarrow X>Y, Z\;\mbox{\sl is}\;X-Y, Y1\;\mbox{\sl is}\;Y+1, q^{\$1>\$2,\$2 \leq 0}(Z,Y1)\\
&& q^{\$1>\$2,\$2\leq 0}(X,Y) \leftarrow X>Y, Z\;\mbox{\sl is}\;X-Y, Y1\;\mbox{\sl is}\;Y+1, q^{\$1\leq \$2,\$2 \leq 0}(Z,Y1)
\end{eqnarray*}

Our assumption is that $\$2 > 0$. The second and the third clauses are 
``irrelevant'' with respect to it. Thus, the only clause that should be analysed is the
first one. The level mapping is thus, redefined as
\[\mid q^{\$1>\$2}(X,Y) \mid\;= c_{\$1>\$2} * \left\{ \begin{array}{ll}
X-Y & \mbox{\rm if}\;\;X>Y \\
0 & \mbox{\rm otherwise}\end{array} \right. +
c_{\$2 > 0} * \left\{ \begin{array}{ll}
Y & \mbox{\rm if}\;\;Y>0 \\
0 & \mbox{\rm otherwise}\end{array} \right.\]

Acceptability decreases imply that 
\[c_{\$1>\$2} (X-Y) + c_{\$2 > 0} Y > c_{\$1>\$2} ((X-Y) - Y) + c_{\$2 > 0} (Y+1),\] i.e., $c_{\$1>\$2} Y > c_{\$2 > 0}$.
Since $Y$ is assumed to be positive this can be
satisfied by taking $c_{\$1>\$2} = 1$ and $c_{\$2 > 0} = 0$. 
This solution does not impose
additional constraints on integer variables. Thus, the analysis terminates
reporting $\$1\leq \$2 \vee (\$1>\$2 \wedge \$2>0)$ as a termination condition.
$\hfill\Box$\end{example}

In order to prove correctness of this algorithm we have to prove its 
termination and partial correctness, i.e., that the symbolic condition
constructed is a termination condition. Termination of the algorithm follows
from termination of its steps discussed earlier and from the finiteness of
the number of integer variables, restricting a number of possible jumps.
Partial correctness follows from the correctness of transformations
and the corresponding result of~\cite{Decorte:DeSchreye:Vandecasteele}.
Observe that after removing ``irrelevant clauses'' 
the non-terminating (for some query) program might
became terminating for it. However, this transformation expresses
the meaning of termination condition: if $c$ holds, clauses defining
predicates such that their adornments are inconsistent with $c$ can not be
unified with $Q$.
Observe also, that at the first traversal of the algorithm 
($c = \mbox{\sl true}$), if $P$ does not have an unreachable code,
this step does not change $P^a$.

\section{Further extensions}
In this section we discuss possible extensions of the algorithm presented 
above. First of all we re-consider inference of adornments, then we
discuss integrating termination analysis of numerical and symbolic 
computations. 
\subsection{Once more about the inference of adornments}
The set of adornments ${\cal A}_p$, inferred in 
Subsection~\ref{subsection:inferring} may sometimes be to week
for inferring precise termination conditions, 
as the following example illustrates.
\begin{example}
\label{example:permuted}
Consider the following program:
\begin{eqnarray*}
&& p(X,Y) \leftarrow X < 0, Y1\;\mbox{\sl is}\;Y+1, p(Y1,X).
\end{eqnarray*}
The maximal prefix of the rule above is $X < 0$, thus,
${\cal A}_p = \{\$1 < 0, \$1 \geq 0\}$. The only termination condition
that will be found is $\$1 \geq 0$, while the precise termination
condition is $\$1 \geq 0 \vee (\$1 < 0 \wedge \$2\geq -1)$.
$\hfill\Box$\end{example}

The problem occured due to the fact that ${\cal A}_p$ restricts 
only {\em some} subset of integer argument positions, while 
for the termination proof information on integer arguments outside of 
this subset may be needed. Thus, we need to infer information
on some variables, given some information on some other variables.
\eat{
This is to be done by means of interargument relations with an exception
of the built-in predicates. For them interargument relations follow
from the predefined semantics, for example the only valid interargument
relation for $\mbox{\sl is}/2$ is $\{(m,n)\mid \mbox{\rm numerical values
of $m$ and $n$ are equal}\}$ and for $>/2$ it is 
$\{(m,n)\mid \mbox{\rm numerical value of $m$ is greater than of $n$}\}$. 
}
\begin{definition}
\label{def:extention}
Let $P$ be a program, let $p$ be a predicate in $P$, let
$C_q$ be a set of symbolic constraints over the integer 
argument positions of $q$, and $C = \cup_{q\in P} C_q$.
A symbolic constraint $c$ over the integer argument positions 
of $p$ is called an 
{\em extension of $C$} if exists $r\in P$, defining $p$,
such that
some integer argument position denominator appearing in $c$
does not appear in $C_p$, and
$c$ is implied by some $c_q\in C_q$ for the 
recursive subgoals and some interargument relations for the 
non-recursive ones.
\end{definition}
Let $C$ be a set of symbolic constraints over the integer 
argument positions of $p$ and let $\varphi(C)$ be
$C \cup \{c\mid c\;\mbox{\rm is an extension of $C$}\}$. 
Define the set of adornments for $p$ as
$\{c'_1\wedge\ldots\wedge c'_n\mid c'_i\in \varphi^{*}({\cal C}_p)\;\mbox{\rm or}\;\neg c'_i\in \varphi^{*}({\cal C}_p)\}$, where $\varphi^{*}(C)$ is a 
fixpoint of powers of $\varphi$ and ${\cal C}_p$ is defined as in Subsection~\ref{subsection:inferring}.

\begin{example}
Consider once more Example~\ref{example:permuted}. Symbolic comparison
$\$1 < 0 \wedge \$2 < -1$ is the only extension of ${\cal C}_p = \{\$1 < 0\}$,
\eat{
.
, since it contains a new integer argument position denominator that does 
not appear in ${\cal C}_p$, i.e., $\$2$, and since it is implied
by taking $\$1 < 0$ for the recursive subgoal $p(Y1,X)$ and
interargument relations for $X>0$ and $Y1\;\mbox{\sl is}\;Y+1$.
This is the only extension of ${\cal C}_p$}
i.e.,
$\varphi({\cal C}_p) = \{\$1 < 0, \$1 < 0 \wedge \$2 < -1\}$. 
All integer argument positions denominators already
appear in $\varphi({\cal C}_p)$. Thus, $\varphi^{*}({\cal C}_p) = \varphi({\cal C}_p)$ and the set of adornments 
is $\{\$1<0 \wedge\$2<-1, (\$1<0 \wedge \$2\geq -1) \vee \$1\geq 0\}$. 
$\hfill\Box$\end{example}

An alternative approach to propagating such an information 
was suggested in~\cite{Dershowitz:Lindenstrauss:Sagiv:Serebrenik}. 
To capture interaction between the variables
a graph was constructed with integer argument positions as vertices and
a ``can influence''-relation as edges. This allows to propagate the
existing adornments but not to infer the new ones and thus, is less 
precise than the approach presented in this subsection.

\eat{
\subsection{Towards general term orderings}
Instead of basing our research on the previous results on acceptability
we could use similar results on term-acceptability~\cite{Serebrenik:DeSchreye:LOPSTR2001:pre}, i.e., acceptability with respect to well-founded term 
orderings. 
This approach allowed to analyse examples such as {\sl derivative}~\cite{DM79:cacm}, {\sl distributive law}~\cite{Dershowitz:Hoot}, {\sl boolean ring}~\cite{Hsiang} and others.

One of the general term orderings that can be extremely useful is
the lexicographic ordering on the set of strings. 
In our case instead of strings vectors of primitive
level mappings are used. 
\begin{example}
This program computes a function, similar
to the Ackermann's function.
\begin{eqnarray*}
&& \mbox{\sl sack}(10,N,N1) \leftarrow N1\;\mbox{\sl is}\;N+1.\\
&& \mbox{\sl sack}(M1,10,V)\leftarrow M1 < 10, M\;\mbox{\sl is}\;M1+1, \mbox{\sl sack}(M,11,V).\\
&& \mbox{\sl sack}(M1,N1,V)\leftarrow M1 < 10, N1 < 10, N\;\mbox{\sl is}\;N1+1, \mbox{\sl sack}(M1,N,V1),\\
&& \hspace{2.9cm} M\;\mbox{\rm is}\;M1+1, \mbox{\sl sack}(M,V1,V).
\end{eqnarray*}
The transformation starts as usual and the following adornments are inferred
$\{((\$1 < 10 \wedge \$2 > 10) \vee \$1 > 10),(\$1 < 10 \wedge \$2 < 10),
(\$1 < 10 \wedge \$2 = 10),\$1 \geq 10\}$. From the clauses obtained
after adorning two are recursive. For the sake of simplicity we discuss
only one of them, the second one is analysed analogously.
\begin{eqnarray*}
&& \mbox{\sl sack}^{\$1 < 10 \wedge \$2 < 10}(M1,N1,V)\leftarrow M1 < 10, N1 < 10, N\;\mbox{\rm is}\;N1+1, \\ 
&& \hspace{1.5cm} \mbox{\sl sack}^{\$1 < 10 \wedge \$2 = 10}(M1,N,V1), M\;\mbox{\rm is}\;M1+1, \mbox{\sl sack}^{\$1 < 10 \wedge \$2 < 10}(M,V1,V).
\end{eqnarray*}
Let $\succ$ be defined as
$\mbox{\sl sack}^{\$1 < 10 \wedge \$2 < 10}(m_1,n_1,v_1)\;\succ\;
\mbox{\sl sack}^{\$1 < 10 \wedge \$2 < 10}(m_2,n_2,v_2)$, 
if either $10 - m_1 > 10 - m_2$ or $(10 - m_1 = 10 - m_2) \wedge
(10 - n_1 > 10 - n_2)$. 
The ordering is well-founded by the well-foundedness of the naturals.
One can easily see that the following decreases hold, implying termination:
\begin{eqnarray*}
&& \mbox{\sl sack}^{\$1 < 10 \wedge \$2 < 10}(M1,N1,V) \;\succ\;
   \mbox{\sl sack}^{\$1 < 10 \wedge \$2 < 10}(M1,N,V1),\;\mbox{\rm where $N = N1+1$}\\
&& \mbox{\sl sack}^{\$1 < 10 \wedge \$2 < 10}(M1,N1,V) \;\succ\;
   \mbox{\sl sack}^{\$1 < 10 \wedge \$2 < 10}(M,V1,V),\;\mbox{\rm where $M = M1+1$}\;\Box
\end{eqnarray*}
$\hfill\Box$\end{example}
} 

\subsection{Integrating numerical and symbolic computation}
As already claimed in the Introduction numerical computations
form an essential part of almost any real-world program. Sometimes, 
such computation is ``pure numerical'', that is does not involve
any reasoning on the symbolic level, such as in the examples above.
However, sometimes numerical computation is interleaved with a symbolic
one as illustrated by the following example, collecting leaves of a tree with
a variable branching factor, being a common data structure
in natural language processing~\cite{Pollard:Sag}.
\begin{example}
\label{example:collect}
\begin{eqnarray*}
&& \mbox{\sl collect}(X,[X|L],L)\leftarrow \mbox{\sl atomic}(X).\\
&& \mbox{\sl collect}(T,L0,L) \leftarrow \mbox{\sl compound}(T), \mbox{\sl functor}(T,\_,A), \\ 
&& \hspace{1.0cm} \mbox{\sl process}(T, 0, A, L0, L).\\
&& \mbox{\sl process}(\_,A,A,L,L).\\
&& \mbox{\sl process}(T,I,A,L0,L2) \leftarrow
	I < A,
	I1\;\mbox{\sl is}\;I+1,
	\mbox{\sl arg}(I1, T, \mbox{\sl Arg}),\\
&& \hspace{1.0cm} \mbox{\sl collect}(\mbox{\sl Arg}, L0,L1),
	\mbox{\sl process}(T,I1,A,L1,L2).\nonumber
\end{eqnarray*}
To prove termination of queries $\{\mbox{\sl collect}(t, v, [])\}$,
where $t$ is a tree and $v$ is a  variable, 
the following three decreases should be shown: between a call to
{\sl collect} and a call to {\sl process} in the second clause,
between a call to {\sl process} and a call to {\sl collect} in the fourth one
and between two calls to {\sl process} in the same clause. The first
and the second decreases
 can be shown only by a symbolic level mapping, the third
one---only by the numerical approach. 
$\hfill\Box$\end{example}

Thus, our goal is to {\sl combine} the existing symbolic approaches with
the numerical one presented so far. One of the possible ways to do so is
to combine two level mappings, $\mid\cdot\mid_1$ and $\mid\cdot\mid_2$, for
example, 
by mapping each atom $A\in B^E_P$ to a pair of natural numbers
$(\mid\!A\mid_1,\mid\!A\mid_2)$.
Then an ordering relation on the atoms can be defined, based on the 
lexicographic ordering of such pairs. Well-foundedness of the natural
numbers implies that this ordering is well-founded and the framework of 
term-acceptability~\cite{Serebrenik:DeSchreye:LOPSTR2001:pre}, allows to
reason on termination of programs in terms of decreases on such orderings.

\begin{example}
Continue Example~\ref{example:collect}. Let $\varphi: B^E_P \rightarrow
({\cal N} \cup {\cal N}^2)$ be a following
mapping: $\varphi(\mbox{\sl collect}(t,l0,l)) = \|t\|$, 
$\varphi(\mbox{\sl process}(t,i,a,l0,l))  = (\|t\|, a-i)$
where $\|\cdot\|$ is a term-size norm. Then, the three decreases
are satisfied with respect to $>$,
such that $A_1 > A_2$ if and only if $\varphi(A_1) \succ \varphi(A_2)$, where
$\succ$ is the lexicographic ordering on ${\cal N} \cup {\cal N}^2$.
$\hfill\Box$\end{example}

This integrated approach allows to analyse correctly examples such as
{\sl ground}, {\sl unify}, {\sl numbervars}~\cite{Sterling:Shapiro} and
Example 6.12 in~\cite{Dershowitz:Lindenstrauss:Sagiv:Serebrenik}. Moreover,
some numerical examples, such as Ackermann's function, 
that cannot be analysed by extending~\cite{Decorte:DeSchreye:Vandecasteele}
due to the limitations of level mappings defined as linear combinations, 
can be analysed by the integrated approach.

\section{Conclusion}
Termination of numerical computations was studied by a number of 
authors~\cite{Apt:Book,Apt:Marchiori:Palamidessi,%
Dershowitz:Lindenstrauss:Sagiv:Serebrenik}. 
Apt {\em et al.}~\cite{Apt:Marchiori:Palamidessi} provided a declarative
semantics, so called $\Theta$-semantics, for Prolog programs with first-order
built-in, including arithmetic operations. 
In this framework the property of
strong termination, i.e., finiteness of all LD-trees for all possible goals,
was completely characterised based on appropriately tuned notion of 
acceptability. This approach provides important theoretical results, but
seems to be difficult to integrate in automatic tools. In~\cite{Apt:Book}
it 
is claimed that an 
unchanged acceptability condition can be applied to programs
in pure Prolog with arithmetic by defining the level mappings on ground atoms
with the arithmetic relation to be zero. This approach ignores
the actual computation, and thus, its applicability is restricted
to programs using some arithmetics but not really relaying on them, 
such as {\sl quicksort}. Moreover, as
Example~\ref{example:gcd:simplified:2} 
illustrates there are many programs that terminate
only for {\em some\/} queries. Alternatively,
Dershowitz {\em et al.}~\cite{Dershowitz:Lindenstrauss:Sagiv:Serebrenik}
extended the query-mapping pairs 
formalism of~\cite{Lindenstrauss:Sagiv} to deal
with numerical computations. However, this approach inherited the disadvantages
of~\cite{Lindenstrauss:Sagiv}, such as high computational price, as well. 

More research has been done on termination analysis for constraint logic
programs~\cite{Colussi:Marchiori:Marchiori,Mesnard,Ruggieri:CLP,%
Ruggieri:thesis}. Since numerical computations in Prolog should be written
in a way that allows a system to verify their satisfiability we can see 
numerical computations of Prolog as an {\em ideal constraint system}. Thus,
all the results obtained for ideal constraints systems can be applied. 
Unfortunately, the research was either oriented towards theoretical 
characterisations~\cite{Ruggieri:CLP,Ruggieri:thesis} or restricted to
domains isomorphic to ${\cal N}$, such as trees and terms~\cite{Mesnard}.

In a contrast to the approach 
of~\cite{Dershowitz:Lindenstrauss:Sagiv:Serebrenik} that was
restricted to verifying termination,
we presented a methodology for {\sl inferring} 
termination conditions. It is not clear 
whether and how~\cite{Dershowitz:Lindenstrauss:Sagiv:Serebrenik}
can be extended to infer such conditions.
 A main contribution of
this work to the theoretical understanding of termination of numerical 
computations is in situating them in the well-known framework of 
acceptability and allowing integration with the existing approaches
to termination of symbolic computations. The methodology presented
can be integrated in automatic termination analysers, such 
as~\cite{Decorte:DeSchreye:Vandecasteele}. 

The kernel technique is powerful enough to analyse correctly examples such as
{\sl gcd} and {\sl mod}~\cite{Dershowitz:Lindenstrauss:Sagiv:Serebrenik},
all examples appearing in the dedicated to arithmetic Chapter 8 
of~\cite{Sterling:Shapiro}, also being a superset of arithmetical
examples appearing in~\cite{Apt:Book}.
Moreover, our approach gains its power from the underlying
framework of~\cite{Decorte:DeSchreye:Vandecasteele} and thus,
allows to prove termination of some
examples that cannot be analysed correctly by~\cite{Dershowitz:Lindenstrauss:Sagiv:Serebrenik}, similar to {\sl confused delete}~\cite{DeSchreye:Decorte:NeverEndingStory,Decorte:DeSchreye:Vandecasteele}. The extended technique, presented in Section 5, 
allows to analyse correctly examples such as Ackermann's function,
{\sl ground}, {\sl unify}, {\sl numbervars}~\cite{Sterling:Shapiro} and
Example 6.12 in~\cite{Dershowitz:Lindenstrauss:Sagiv:Serebrenik}.

\eat{
Due to the space restrictions we do not present here 
further extensions of our technique (see~\cite{Serebrenik:DeSchreye:cw308}), 
such as more refined domain inference, including propagating information
from one subset of integer argument positions to another one, and integration 
within the framework of 
acceptability based on general term orderings~\cite{Serebrenik:DeSchreye:LOPSTR2001:pre}. The later extension allows to prove termination  
{\em Ackermann's function}. }
 
As a future work we consider a complete implementation of the algorithm.
Due to the use of the constraint solving
techniques we expect it both to be powerful and highly efficient.
\eat{
\section{Acknowledgement}
Alexander Serebrenik is supported by GOA: ``${LP}^{+}$: a second generation
logic programming language''. 
}
\bibliography{/home/alexande/M.Sc.Thesis/main}

\begin{thebibliography}{10}

\bibitem{Apt:Book}
K.~R. Apt.
\newblock {\em From Logic Programming to Prolog}.
\newblock Prentice-Hall Int. Series in Computer Science. Prentice Hall, 1997.

\bibitem{Apt:Marchiori:Palamidessi}
K.~R. Apt, E.~Marchiori, and C.~Palamidessi.
\newblock A declarative approach for first-order built-in's in prolog.
\newblock {\em Applicable Algebra in Engineering, Communication and
  Computation}, 5(3/4):159--191, 1994.

\bibitem{Bossi:Cocco:Basic}
A.~Bossi and N.~Cocco.
\newblock Basic transformation operations which preserve computed answer
  substitutions of logic programs.
\newblock {\em J. Logic Programming}, 16:47--87, May 1993.

\bibitem{Bossi:Cocco}
A.~Bossi and N.~Cocco.
\newblock Preserving universal temination through unfold/fold.
\newblock In G.~Levi and M.~Rodr\'{\i}guez-Artalejo, editors, {\em Algebraic
  and Logic Programming}, pages 269--286. Springer Verlag, 1994.
\newblock LNCS 850.

\bibitem{Colussi:Marchiori:Marchiori}
L.~Colussi, E.~Marchiori, and M.~Marchiori.
\newblock On termination of constraint logic programs.
\newblock In U.~Montanari and F.~Rossi, editors, {\em Principles and Practice
  of Constraint Programming - CP'95,}, pages 431--448. Springer Verlag, 1995.
\newblock LNCS 976.

\bibitem{DeSchreye:Decorte:NeverEndingStory}
D.~De~Schreye and S.~Decorte.
\newblock Termination of logic programs: The never-ending story.
\newblock {\em J. Logic Programming}, 19/20:199--260, May/July 1994.

\bibitem{DeSchreye:Verschaetse:Bruynooghe}
D.~De~Schreye, K.~Verschaetse, and M.~Bruynooghe.
\newblock A framework for analyzing the termination of definite logic programs
  with respect to call patterns.
\newblock In I.~Staff, editor, {\em Proc. of the Int. Conf. on Fifth Generation
  Computer Systems.}, pages 481--488. IOS Press, 1992.

\bibitem{Decorte:DeSchreye:98}
S.~Decorte and D.~De~Schreye.
\newblock Termination analysis: some practical properties of the norm and level
  mapping space.
\newblock In J.~Jaffar, editor, {\em Proc. of the 1998 Joint Int. Conf. and
  Symp. on Logic Programming}, pages 235--249. MIT Press, June 1998.

\bibitem{Decorte:DeSchreye:Vandecasteele}
S.~Decorte, D.~De~Schreye, and H.~Vandecasteele.
\newblock Constraint-based termination analysis of logic programs.
\newblock {\em ACM Transactions on Programming Languages and Systems (TOPLAS)},
  21(6):1137--1195, November 1999.

\bibitem{Dershowitz:Lindenstrauss:Sagiv:Serebrenik}
N.~Dershowitz, N.~Lindenstrauss, Y.~Sagiv, and A.~Serebrenik.
\newblock {A} general framework for automatic termination analysis of logic
  programs.
\newblock {\em Appl. Algebr. Eng. Commun. Comput.}, 2001.
\newblock accepted.

\bibitem{Kawamura:Kanamori}
T.~Kawamura and T.~Kanamori.
\newblock Preservation of stronger equivalence in unfold/fold transformation.
\newblock {\em Theoretical {C}omputer {S}cience}, 75(1\&2):139--156, Sept.
  1990.

\bibitem{Lindenstrauss:Sagiv}
N.~Lindenstrauss and Y.~Sagiv.
\newblock Automatic termination analysis of logic programs.
\newblock In L.~Naish, editor, {\em Proc. of the Fourteenth Int. Conf. on Logic
  Programming}, pages 63--77. MIT Press, July 1997.

\bibitem{Lindenstrauss:Sagiv:Serebrenik:L}
N.~Lindenstrauss, Y.~Sagiv, and A.~Serebrenik.
\newblock Unfolding mystery of the {\em mergesort\/}.
\newblock In N.~Fuchs, editor, {\em Proc. of the Seventh Int. Workshop on Logic
  Program Synthesis and Transformation}. Springer Verlag, 1998.
\newblock LNCS 1463.

\bibitem{Mesnard}
F.~Mesnard.
\newblock Inferring left-terminating classes of queries for constraint logic
  programs.
\newblock In M.~Maher, editor, {\em Proc. {JICSLP'96}}, pages 7--21. The {MIT}
  {P}ress, 1996.

\bibitem{Pollard:Sag}
C.~Pollard and I.~A. Sag.
\newblock {\em Head-driven Phrase Structure Grammar}.
\newblock The University of Chicago Press, 1994.

\bibitem{Ruggieri:CLP}
S.~Ruggieri.
\newblock Termination of constraint logic programs.
\newblock In P.~Degano, R.~Gorrieri, and A.~Marchetti-Spaccamela, editors, {\em
  Automata, Languages and Programming, 24th International Colloquium,
  ICALP'97}, pages 838--848. Springer Verlag, 1997.
\newblock LNCS 1256.

\bibitem{Ruggieri:thesis}
S.~Ruggieri.
\newblock {\em Verification and validation of logic programs}.
\newblock PhD thesis, Universit{\'a} di Pisa, 1999.

\bibitem{Serebrenik:DeSchreye:LOPSTR2001:pre}
A.~Serebrenik and D.~De~Schreye.
\newblock {N}on-transformational termination analysis of logic programs, based
  on general term-orderings.
\newblock In K.-K. Lau, editor, {\em Pre-Proceedings of Tenth International
  Workshop on Logic-based Program Synthesis and Transformation, 2000}, pages
  45--54. University of Manchester, 2000.
\newblock Department of Computer Science, Univ. of Manchester, ISSN 1361-6161.
  Report number UMCS-00-6-1, URL :
  http://www.cs.man.ac.uk/cstechrep/titles00.html.

\bibitem{Sterling:Shapiro}
L.~Sterling and E.~Shapiro.
\newblock {\em The {A}rt of {P}rolog}.
\newblock The {MIT} Press, 1994.

\end{thebibliography}
\bibliographystyle{abbrv}
\end{document}